\documentclass[11pt,a4paper]{article}

\usepackage{latexsym}
\usepackage{amssymb}
\usepackage{amsmath}
\usepackage{amsfonts}
\usepackage{mathrsfs}
\usepackage{cite}
\usepackage{color}
\usepackage{bm}

\catcode`@=11
\renewcommand{\section}{\@startsection{section}{1}{0pt}{\medskipamount}
    {\medskipamount}{\large\bf}} \numberwithin{equation}{section}
\catcode`@=12

\newcommand{\be}{\begin{equation}}
\newcommand{\ee}{\end{equation}}

\def\tr{{\rm tr}}
\def\Tr{{\rm Tr}}
\def\cN{{\cal N}}
\def\cD{{\cal D}}

\def\bea{\begin{eqnarray}}
\def\eea{\end{eqnarray}}
\def\nn{\nonumber}

\def\cN{{\cal N}}
\def\cD{{\cal D}}

\def\f{\frac}

\def\nn{\nonumber}

\def\d{\delta}

\def\sB{\stackrel{\frown}{\square}}
\def\bsB{\stackrel{\frown}{\bm{\square}}}

\def\eq{\eqref}
\def\pr{\partial}
\def\nb{\nabla}

\numberwithin{equation}{section}

\begin{document}


\begin{center}
\vspace{1cm}

{\Large \bf The renormalization structure of $6D,\, \cN=(1,0)$
\vspace*{1mm}

supersymmetric higher-derivative gauge theory}
%

\vspace{0.7cm}

{\bf
 I.L. Buchbinder\footnote{joseph@tspu.edu.ru }$^{\,a,b,c}$,
 E.A. Ivanov\footnote{eivanov@theor.jinr.ru}$^{\,c,d}$,
 B.S. Merzlikin\footnote{merzlikin@tspu.edu.ru}$^{\,e,a,c,}$,
 K.V. Stepanyantz\footnote{stepan@m9com.ru}$^{\,f,c}$
}
 \vspace{0.7cm}

{\it
 $^a$ Department of Theoretical Physics, Tomsk State Pedagogical University,\\ 634061, Tomsk,  Russia \\ \vskip 0.15cm
 $^b$ National Research Tomsk State University, 634050, Tomsk, Russia \\ \vskip 0.1cm
 $^c$ Bogoliubov Laboratory of Theoretical Physics, JINR, 141980 Dubna,
 Moscow region, Russia \\ \vskip 0.1cm
 $^d$ Moscow Institute of Physics and Technology, 141700 Dolgoprudny, Moscow Region, Russia\\ \vskip 0.1cm
 $^e$ Tomsk State University of
 Control Systems and Radioelectronics, 634050 Tomsk, Russia\\ \vskip 0.1cm
 $^f$ Department of Theoretical Physics, Moscow State University, 119991, Moscow, Russia
}
\end{center}

\vspace{0.1cm}

\begin{abstract}
We consider the harmonic superspace formulation of higher-derivative
$6D, \cN=(1,0)$ supersymmetric gauge theory and its minimal coupling to a hypermultiplet. In components, the kinetic term for the
gauge field in such a theory involves four space-time derivatives.
The theory is quantized in the framework of the superfield
background method ensuring manifest $6D, \,\cN=(1,0)$ supersymmetry
and the classical gauge invariance of the quantum effective action.
We evaluate the superficial degree of divergence and prove it to be
independent of the number of loops. Using the regularization by
dimensional reduction, we find possible counterterms and show that
they can be removed by the coupling constant renormalization for any
number of loops, while the divergences in the hypermultiplet sector
are absent at all. Assuming that the deviation of the gauge-fixing
term from that in the Feynman gauge is small, we explicitly
calculate the divergent part of the one-loop effective action in the
lowest order in this deviation.
In the approximation considered, the result  is independent of the gauge-fixing parameter and agrees with
the earlier calculation for the theory without a hypermultiplet.
\end{abstract}


\setcounter{footnote}{0} \setcounter{page}{1}

\section{Introduction}

The $6D, \cN=(1,0)$ supersymmetric higher-derivative
gauge theory was firstly constructed in \cite{Ivanov:2005qf}, starting from its harmonic-superspace formulation.
It describes a self-interacting non-abelian gauge multiplet with the kinetic term of the fourth order in derivatives.

The ordinary $\cN=(1,0)$ supersymmetric Yang-Mills
(SYM) theory in $6D$ has a dimensionful coupling constant and for this reason  is non-renormalizable. The UV behavior of such a theory
was studied by direct quantum calculations in the component approach
\cite{Fradkin:1982kf,Marcus:1983bd,Marcus:1984ei,Kazakov:2002jd,Buchbinder:2019cnh},
by using the gauge and supersymmetry methods
\cite{Howe:1983jm,Howe:2002ui,Bossard:2009sy,Bossard:2009mn,Bossard:2015dva},
by applying the background field method in superspace
\cite{Buchbinder:2016url,Buchbinder:2017ozh} and, recently,  by the modern amplitude
techniques (see, e.g., \cite{Bork:2015zaa} and references therein). In contrast to the
gauge theory with the standard kinetic term the higher-derivative model
with four space-time derivatives \cite{Ivanov:2005qf}
possesses  a dimensionless coupling constant. It is renormalizable by power
counting and conformally invariant \cite{Ivanov:2005kz} at the
classical level, although reveals the quantum conformal anomaly
\cite{Smilga:2004cy,Smilga:2006ax}. The one-loop beta function for this theory has been calculated
in \cite{Ivanov:2005qf,Casarin:2019aqw} using the component formulation and, recently,
by applying the supergraph analysis in harmonic superspace
\cite{Buchbinder:2020tnc}.

Although the higher-derivative models are plagued by the ghost
states in the spectrum, they are widely used in classical and
quantum field theory\footnote{In fact, there were numerous  attempts
to show that in the interacting higher-derivative theories (with or
without supersymmetry) the ghosts can be arranged in such a way that
they do not contribute to the observables (see, e.g.,
\cite{Hawking:1985gh,Smilga:2004cy,Robert:2006nj,{Smilga:2017arl},Anselmi:2020tqo}
and the references therein).}. The main attractive feature of such
models is seen in a possibility to improve their ultraviolet
behavior as compared to the corresponding conventional theories. In
particular, the inclusion of the four-derivative terms into the
general relativity Lagrangian yields a renormalizable quantum
gravity with matter (see, e.g.,
\cite{Fradkin:1981iu,Buchbinder:1989jd,Buchbinder:1992rb}). The
higher derivatives appear naturally (and inevitably) in such field
models as conformal (super)gravities and (super)conformal
higher-spin theories (see, e.g.,
\cite{Fradkin:1985am,Segal:2002gd,Grigoriev:2016bzl,Adamo:2018srx,Buchbinder:2019yhl,Kuzenko:2019eni}
and the references therein). In this paper we investigate the
quantum aspects of the six-dimensional supersymmetric
higher-derivative theory pioneered in \cite{Ivanov:2005qf}.

Our aim here is to study the divergence structure of
the theory under consideration in a manifestly supersymmetric and
covariant way on the basis of the background superfield method in
$6D, \,\cN=(1,0)$ harmonic superspace \cite{Howe:1985ar,Zupnik:1986da}.
The general concept of the harmonic superspace has been introduced in
\cite{Galperin:1984av} (see also the book \cite{Galperin:2001uw}).
The background field method in $6D, {\cal N}=(1,0)$
harmonic superspace was formulated in our works
\cite{Buchbinder:2016url,Buchbinder:2017ozh}\footnote{$6D, {\cal
N}=(1,0)$ background superfield method is in many aspects similar  to the
one developed for $4D, \,{\cal N}=2$ SYM theory
\cite{Buchbinder:1997ya,Buchbinder:1998np}.}. Taking into account
the structure of the superfield propagators and vertices, as well as
the manifest supersymmetry and gauge invariance of the effective action,
we calculate the superficial degree of the divergence and prove
that it does not depend on the number of loops. In the case of using the regularization by dimensional reduction the power counting implies that
any possible counterterm is proportional to the classical
higher-derivative action. This means that the theory under consideration is
multiplicatively renormalizable.

The basic feature of the background field method is the use of special gauge conditions involving a dependence on the background
fields. However, due to the presence of higher derivatives, the
previously used background field gauge-fixing conditions are not
convenient in the theory under consideration  and the superfield
gauge-fixing action should be constructed in a different way. In
this paper we introduce the appropriate family of the gauge-fixing
conditions depending on one real (gauge) parameter $\xi_0$. The
value $\xi_0 =1$ corresponds to the minimal gauge. Nevertheless,
keeping $\xi_0$ arbitrary provides us with an efficient tool of
checking the correctness of the calculations.  Indeed, the
divergences of the multiplicatively renormalizable theory in the
background field method should not depend on the gauge choice
\cite{Voronov:1982ph} (see also \cite{Barnich:2000yx} and the references
therein).

It is worth noting that using an arbitrary parameter
$\xi_0$ extremely complicates the calculations of divergences, since
the background-field dependent differential operator in the
quadratic part of the action becomes non-minimal\footnote{The
higher-derivative differential operator acting on space-time fields is called non-minimal
if the higher derivatives are not assembled into powers of d'Alambertian.}. The
most powerful manifestly covariant method to work with such
operators is the generalized proper-time technique
\cite{Barvinsky:1985an}, which can in principe be reformulated for
superfield theories. However, in some cases the calculations
can be further simplified, assuming that the deviation of the gauge-fixing
parameter from its value in the minimal gauge is small and the
calculations are performed to the lowest order in this small
parameter. In this paper we follow just this strategy. We assume
that the value of the gauge parameter $\xi_0$ is close to 1 and
calculate the one-loop divergences in the lowest order in the
deviation $(\xi_0-1)$. Then we explicitly demonstrate that  the
logarithmic divergencies are gauge independent in the considered
approximation, as is expected for a multiplicatively renormalizable
theory. The numerical coefficient before the one-loop beta function
precisely matches the result of refs. \cite{Ivanov:2005qf,Casarin:2019aqw,Buchbinder:2020tnc}.

Also we study the UV behavior for the model in which the higher-derivative gauge multiplet is minimally
coupled to the hypermultiplet with the
standard kinetic term (yielding the standard second and first-order kinetic terms for
the relevant physical bosonic and fermionic fields). We demonstrate that the presence of such
a hypermultiplet does not destroy the
renormalizability of the theory (regularized by dimensional reduction) and that no divergent contributions
depending on the background hypermultiplet appear even in non-minimal
gauges. At the one-loop level, the only  effect of adding the hypermultiplet is the change of the absolute
value of the coefficient before the beta function in the gauge superfield sector.

The paper is organized as follows. Section \ref{Section_Classical}
presents the classical formulation of the theories under
consideration. In Section \ref{Section_Quantization} we perform the
quantization of the theory with higher derivatives in the harmonic
$6D,\, {\cal N}=(1,0)$ superspace, based  upon the background superfield method.
In Section \ref{Section_Index} we discuss the general structure of divergencies in the theory
and calculate the relevant divergence degree. In Section
\ref{Section_Divergences} we find the divergent contribution to
one-loop effective action for the higher-derivative gauge theory.
The UV properties of the higher-derivative gauge theory coupled
to a hypermultiplet are explored in Section \ref{Section_Hypermultiplet}. The concluding Section
\ref{Section_Conclussion} contains a brief summary of our results
and a list of possible directions of the future work.

\section{Superfield formulation of $6D,\, \cN=(1,0)$ higher-derivative gauge theory}
\label{Section_Classical}

The classical action of the higher-derivative $6D, \,\cN=(1,0)$ supersymmetric gauge theory constructed
in \cite{Ivanov:2005qf} is written in the harmonic superspace as
 \bea
 S_0=\pm \f1{2g_0^2}\tr \int d\zeta^{(-4)} du\, (F^{++})^2\,, \label{S0}
 \eea
where $g_0$ is a dimensionless coupling constant.
The integration measure over analytic subspace in harmonic superspace is denoted by
$d\zeta^{(-4)} = d^6 x_{(\rm an)}\, (D^-)^4$, see \cite{Galperin:2001uw}
for details. The covariant strength of the analytic gauge
superfield $V^{++}$ is defined by the expression
 \bea
 F^{++} = (D^+)^4 V^{--} = -\frac{1}{24}\varepsilon^{abcd}
 D^+_a D^+_b D^+_c D^+_d V^{--},
 \eea
where
 \bea
\label{V--} V^{--}(z,u)=\sum\limits^{\infty}_{n=1} (-i)^{n+1} \int
du_1\ldots du_n\, \frac{V^{++}(z,u_1 )\ldots V^{++}(z,u_n )}{(u^+
u^+_1)(u^+_1 u^+_2)\ldots (u^+_n u^+)}
 \eea
is a non-analytic superfield introduced in \cite{Zupnik:1986da}. As we noted in \cite{Buchbinder:2020tnc}, the
overall sign of the action in higher derivative theories cannot be
fixed from the standard requirement of the positive energy.

The action \eq{S0} is invariant under the gauge transformation
 \bea\label{tr1}
 \d_\lambda V^{++} = -D^{++} \lambda -
 i [V^{++}, \lambda]\,,\qquad \d_\lambda F^{++} = i[\lambda,F^{++}]\,,
 \label{gtrans}
 \eea
with the Hermitian analytic superfield parameter $\lambda$ taking values in the Lie algebra of the gauge group.
In our notation the generators are normalized by the condition $\tr(
t^I t^J )= \tfrac12 \d^{IJ}$ and satisfy the commutation relation
$[t^I,t^J]=if^{IJK}t^K$.

The component structure of the action \eq{S0} was discussed in
\cite{Ivanov:2005qf}. The $6D$ gauge multiplet $V^{++}$
in the Wess-Zumino gauge involves\footnote{Hereafter we denote the space-time indices as $
M,N = 0,\ldots,5$,  the spinor ones as $a,b=1,\ldots,4$ and the indices of
$SU(2)$ R-symmetry group as $i,j =1,2$.} the vector field $A_M$, the  Weyl
fermion $\psi^{a i }$ and an $SU(2)$ triplet of scalar fields
$\cD^{(ij)}$. In the component form the action \eq{S0} contains
four derivatives in the kinetic term for the gauge field $A_M$ and three
derivatives in the kinetic term for the gaugino $\psi^{ai}$. The scalar
fields $\cD^{ij}$ are also dynamical, with standard two derivatives
in the kinetic term.

The higher-derivative gauge theory \eq{S0} in interaction with
the hypermultiplet $q^{+}$ possessing  the standard kinetic term (see
\cite{Galperin:2001uw} for details) is described by the action
 \bea
 \widetilde S_0 = S_0
 - \int d\zeta^{(-4)} du\, \widetilde q^{+}\nb^{++}q^+\,. \label{S02}
 \eea
The hypermultiplet can be placed in an arbitrary representation $R$ of
the gauge group. Correspondingly,  the harmonic covariant derivative in \eq{S02}, $\nb^{++} = D^{++} + i V^{++}$,
is defined in this representation.  The action \eq{S02} is invariant under the
transformation $\eq{tr1}$ supplemented by the transformation of the
hypermultiplet,
 \bea
\d_{\lambda}q^{+} = i\lambda q^{+}\,.
 \eea

The classical equations of motion for the model \eq{S02} read
 \bea
\f{\delta \widetilde S_0}{\delta V^{++I}} &=& \pm \f{1}{2g_0^2} (\sB
F^{++})^I
- i \widetilde q^{+}\, T^I\, q^+ = 0\,, \nn \\
\f{\delta \widetilde S_0}{\delta \widetilde q^{+}} &=& -\nb^{++} q^+=0\,,
 \eea
where $T^I,\, I=1,..,{\rm dim}\, G$ are the gauge group  generators
in the representation $R$. The operator
\begin{equation}\label{smile}
\sB=\frac{1}{2}(D^+)^4(\nb^{--})^2
\end{equation}
acting on a space of analytic superfields is reduced to the
covariant superfield d'Alembertian
\begin{eqnarray}
\label{Box_First_Part} \sB = \eta^{MN} \nabla_M \nabla_N + i W^{+a}
\nabla^{-}_a + i F^{++} \nabla^{--} - \frac{i}{2}(\nabla^{--} F^{++}),
\end{eqnarray}
where $\eta_{MN},$ is $6D$ Minkowski metric (with the signature $(+ - - -)$) and the covariant
derivatives are defined by the relations \bea \nb^{--}=D^{--} + i
V^{--},\qquad [\nabla^{--}, D^+_a] = \nabla^-_a,\qquad
[D^+_a,\nabla^-_b] = i (\gamma^M)_{ab}\nabla_M. \eea We have
introduced in \eq{Box_First_Part} the gauge superfield strength
 \bea
 W^{+a} = -\f16\varepsilon^{abcd} D^+_b D^+_c D^+_d V^{--}\,,
 \eea
such that $F^{++} = \tfrac14 D^+_a W^{+a}$.

\section{Effective action}
\label{Section_Quantization}

For defining the quantum version of the theory \eq{S0} we use the background
superfield method. Following the general procedure, the superfield
$V^{++}$ is split into the sum of the ``background'' superfield
$\bm{V}^{++}$ and the ``quantum'' one $v^{++}\,$,
 \be
 V^{++} = \bm{V}^{++} + v^{++}.
 \ee

Like in \cite{Buchbinder:2016url}, we use the gauge-fixing
function ${\cal F}^{(+4)}_\tau=D^{++}v^{++}_\tau$, where the subscript
$\tau$ means $\tau$-frame which is constructed with the help of the
background bridge superfield. The corresponding action for the real
analytic fermionic Faddeev-Popov ghosts $b$ and $c$ has the same form
as in $6D, \, \cN=(1,0)$ SYM theory with the standard kinetic
term \cite{Buchbinder:2016url,Buchbinder:2017ozh}
 \begin{equation}\label{FP}
S_{\mbox{\scriptsize FP}}[\,b, c, v^{++}, \bm{V}^{++}] =\tr\int d\zeta^{(-4)} du\,
b\,\bm{\nb}^{++}\Big(\bm{\nb}^{++} c + i[v^{++}, c]\Big),
 \end{equation}
where $\bm{\nb}^{\pm\pm} = D^{\pm\pm} + i\bm{V}^{\pm\pm}$.

The effective action $\Gamma[\bm{V}^{++}]$ of the theory is defined as in \cite{Buchbinder:2016url}
 \bea \label{path}
 e^{i\Gamma[\bm{V}^{++}]}=\int {\cal D}v^{++} {\cal D}b\, {\cal D} c\,
 \delta[{\cal F}^{(+4)}-f^{(+4)}]\,
 \exp\Big(i \big\{ S_0[\bm{V}^{++} + v^{++}]
 + S_{\mbox{\scriptsize FP}}[\, b, c, v^{++}, \bm{V}^{++}]\big\}\Big)\,.
 \eea
The superfield $f^{(+4)}(\zeta,u)$ is an external analytic
superfield independent of the background superfield $\bm{V}^{++}$
which takes value in the
Lie algebra of the gauge group. The effective action by construction is
invariant under the background gauge transformations
\begin{equation}\label{cltr}
\delta \bm{V}^{++}=-\bm{\nb}^{++}\lambda, \quad \delta v^{++} = i[\lambda, v^{++}].
\end{equation}

Following \cite{Buchbinder:2016url}, we average the delta-function
$\d(F^{(+4)}-f^{(+4)})$ with the weight
 \be
 \label{weight}
 1=\Delta_{\mbox{\scriptsize NK}}[\bm{V}^{++}]\exp\Big\{\mp \frac{i}{2g_0^2\xi_0}\int d^{14}z du_1\,
du_2\,f_\tau^{(+4)}(u_1)\f{(u^-_1u^-_2)}{(u^+_1u^+_2)^3}\bsB_2
f_\tau^{(+4)}(u_2)\Big\},
 \ee
where $\bsB = \frac{1}{2} (D^+)^4 (\bm{\nb}^{++})^2$ and $\xi_0$ is
an arbitrary real parameter. Note that, as distinct from the theory
without higher derivatives, in \eq{weight} there appears an extra operator $\bsB$. The
factor $\Delta[\bm{V}^{++}]$ yields the Nielson-Kallosh determinant.
It is convenient to present it as the functional integral over the
bosonic real analytic superfield $\varphi$ and anticommuting
analytic superfields $\chi^{(+4)}$ and $\sigma$, all in the adjoint
representation of the gauge group,
 \be
\Delta_{\mbox{\scriptsize NK}}[\bm{V}^{++}]= \int {\cal D}\varphi\,
{\cal D}\chi^{(+4)}\, {\cal D}\sigma
\exp\Big\{ i S_{\mbox{\scriptsize NK}}[\varphi,\bm{V}^{++}]\Big\}\,.
 \ee
Here,
 \be
S_{\mbox{\scriptsize NK}} = \tr \int d\zeta^{(-4)}\, du\, \Big(-\frac{1}{2}\varphi
(\bm{\nb}^{++})^2\varphi +\chi^{(+4)}\bsB \sigma\Big) \label{NK}
 \ee
is the Nielsen--Kallosh ghost action.

The gauge-fixing term obtained as a result of the procedure described above reads
 \bea
S_{\mbox{\scriptsize gf}}[v^{++}, \bm{V}^{++}]
= \mp \frac{1}{2g_0^2\xi_0} \tr \int d^{14}z du_1\, du_2\,
\frac{(u_1^- u_2^-)}{(u_1^+ u_2^+)^3} \big(D^{++} v^{++}_{\tau}\big)_2
\big(\bsB_{\tau} D^{++} v^{++}_{\tau}\big)_1.
 \eea
According to \cite{Buchbinder:2020tnc}, it can be equivalently rewritten in the form
  \bea
&& S_{\mbox{\scriptsize gf}} = \pm \frac{1}{2g_0^2\xi_0}\tr \int d\zeta^{(-4)} du\,
v^{++}
\bsB{}^2 v^{++} \nn \\
&&\qquad\qquad
\mp \frac{1}{2g_0^2\xi_0}\tr \int d^{14}z
\f{du_1du_2}{(u^+_1u^+_2)^2}\, v_{\tau,1}^{++} \Big\{(\bsB v^{++}) + \frac{i}{2}
\bm{\nb}^{--} [\bm{F}^{++},v^{++}] \Big\}_{\tau,2}. \qquad \label{SGF}
 \eea
Adding this expression to that part of the classical action which is quadratic in the quantum gauge superfield and was calculated in \cite{Buchbinder:2020tnc},
 \bea
  && S^{(2)}_0 =
  \pm\f{1}{2g_0^2} \tr \int d^{14}z \f{du_1du_2}{(u^+_1u^+_2)^2}
 \Big\{v_{\tau,1}^{++}(\bsB v^{++})_{\tau,2}   + \frac{i}{2}
v^{++}_{\tau,1}[(\bm{\nb}^{--}\bm{F}^{++}),v^{++}]_{\tau,2} \Big\}\qquad\nn\\
&& \mp\f{i}{4g_0^2} \tr \int d^{14}z \f{du_1 du_2}{(u^+_1u^+_2)^2}
 v^{++}_{\tau,1}[\bm{F}^{++},\bm{\nb}^{--} v^{++}]_{\tau,2},
  \label{S2_Gauge0}
 \eea
\noindent
we obtain
 \bea
  && S^{(2)}_{\mbox{\scriptsize gauge}} =
  \pm \frac{1}{2g_0^2\xi_0}\tr \int d\zeta^{(-4)} du\, v^{++}\bsB{}^2 v^{++} \pm\f{1}{2g_0^2}\Big(1-\f1\xi_0\Big)
 \tr \int d^{14}z \f{du_1du_2}{(u^+_1u^+_2)^2}
 \Big\{v_{\tau,1}^{++}(\bsB v^{++})_{\tau,2} \qquad\nn \\
 &&  + \frac{i}{2}
v^{++}_{\tau,1}[(\bm{\nb}^{--}\bm{F}^{++}),v^{++}]_{\tau,2} \Big\} \mp\f{i}{4g_0^2}  \Big(1+\f1\xi_0\Big)\tr \int d^{14}z \f{du_1 du_2}{(u^+_1u^+_2)^2}
 v^{++}_{\tau,1}[\bm{F}^{++},\bm{\nb}^{--} v^{++}]_{\tau,2}.
  \label{S2_Gauge}
 \eea
The quantum effective action which is gauge invariant and
$\cN=(1,0)$ supersymmetric by construction (for details see refs.
\cite{Buchbinder:2016url,Buchbinder:2017ozh,Buchbinder:2019gfb,Buchbinder:2020tnc}) is given by the expression
 \bea\label{Effective_Action}
 e^{i \Gamma[\bm{V}^{++}]} = \int {\cal
D}v^{++}\,{\cal D}q^+\, {\cal D}b\,{\cal D}c\,{\cal D}\varphi\,
{\cal D}\chi^{(+4)}\, {\cal D}\sigma\,
\exp\Big(iS_{\mbox{\scriptsize total}} - \int d\zeta^{(-4)}\, du\,
\frac{\delta \Gamma[\bm{V}^{++}]}{\delta \bm{V}^{++A}}\,
v^{++A}\Big),\label{path2}
 \eea
where the total action has the form
 \bea
 S_{\mbox{\scriptsize total}} &=& S_{0}[\bm{V}^{++}+v^{++}]
 +S_{\mbox{\scriptsize gf}}[v^{++}, \bm{V}^{++}] +S_{\mbox{\scriptsize gh}}\,,
 \eea
and we have denoted $S_{\mbox{\scriptsize gh}} = S_{\mbox{\scriptsize FP}}[b, c,
v^{++}, \bm{V}^{++}] +S_{\mbox{\scriptsize NK}}[\varphi,
\chi^{(+4)},\sigma, \bm{V}^{++}]$.

The effective action defined in (\ref{path2}) has the structure
$\Gamma[\bm{V}^{++}]=S[\bm{V}^{++}] + \Delta\Gamma[\bm{V}^{++}]$,
where $\Delta\Gamma[\bm{V}^{++}]$ accommodates  all quantum corrections to
the classical action.

Further we will consider the structure of the effective action in the one-loop
approximation. From eq. (\ref{Effective_Action}) we see that the
one-loop contribution to the effective action is given by the
functional integral
 \be
 \exp\big( i\Delta\Gamma^{(1)}[\bm{V}^{++}]\big)
 = \int {\cal D}v^{++}\, {\cal D}b\,{\cal D}c\,{\cal D}\varphi\,
 {\cal D}\chi^{(+4)}\,{\cal D}\sigma\,
\exp\big(iS_{\mbox{\scriptsize total}}^{(2)}
[v^{++}, b, c, \varphi, \chi^{(+4)}, \sigma, \bm{V}^{++}]\big)\,,
 \label{Gamma0}
 \ee
where $S_{\mbox{\scriptsize total}}^{(2)}$ denotes a part of the
total action quadratic in the quantum superfields. To present it in the most convenient form, we integrate by parts with respect to the derivative $\bm{\nabla}^{--}_2$ in the last term of the expression (\ref{S2_Gauge}).
After this, taking into account that $e^{-i\bm{b}} \bm{\nabla}^{--} e^{i\bm{b}} = D^{--}$, we can rewrite $S^{(2)}_{\mbox{\scriptsize total}}$ as
  \bea
  S^{(2)}_{\mbox{\scriptsize total}} &=& \pm \frac{1}{2g_0^2\xi_0}\tr \int d\zeta^{(-4)} du\, v^{++}\bsB{}^2 v^{++}
 \pm\f{1}{2g_0^2}
 \tr \int d^{14}z \f{du_1du_2}{(u^+_1u^+_2)^2}
 \Big\{\Big(1-\f1\xi_0\Big) v_{\tau,1}^{++}(\bsB v^{++})_{\tau,2}
 \nn \\
 && + i v^{++}_{\tau,1}[(\bm{\nb}^{--}\bm{F}^{++}),v^{++}]_{\tau,2} \Big\}
  \mp\f{i}{2g_0^2}  \Big(1+\f1\xi_0\Big)\tr \int d^{14}z du_1 du_2\frac{(u_1^+ u_2^-)}{(u^+_1u^+_2)^3}
 v^{++}_{\tau,1}[\bm{F}^{++}, v^{++}]_{\tau,2}
 \nn \\
 && + \tr \int d\zeta^{(-4)}du\, b\,(\bm{\nb}^{++})^{2} c
  +\tr \int d\zeta^{(-4)}du\,\Big(-\frac{1}{2}\varphi
(\bm{\nb}^{++})^2\varphi +\chi^{(+4)}\bsB \sigma\Big).
  \label{S2}
 \eea
After integration over quantum superfields in the functional
integral \eq{Gamma0} we obtain the one-loop quantum correction to
the effective action as a sum of three terms
\footnote{Obviously, the one-loop quantum correction is the same for the upper and lower signs in the first three terms of
\eq{S2}.}
  \bea
 &&\hspace*{-5mm} \Delta\Gamma^{(1)}[\bm{V}^{++}] =
 \frac{i}{2}\,\mbox{Tr}_{(2,2)}\ln \Big\{ \frac{1}{\xi_0}
(\bsB_1)^2\, (D_1^+)^4\delta^{(-2,2)}(u_1,u_2)
+ \Big(1-\frac{1}{\xi_0}\Big) \frac{(D_1^+)^4 \bsB_2 (D_2^+)^4}{(u_1^+ u_2^+)^2} e^{i\bm{b}_1} e^{-i\bm{b}_2}
\qquad\nonumber\\
&&\hspace*{-5mm} + \frac{(D_1^+)^4 (D_2^+)^4}{(u_1^+ u_2^+)^2} e^{i\bm{b}_1} e^{-i\bm{b}_2}
\Big[\, i (\bm{\nabla}^{--}\bm{F}^{++}) -
\frac{i(u_1^+ u_2^-)}{(u_1^+ u_2^+)} \Big(1+\frac{1}{\xi_0}\Big) \bm{F}^{++} \Big]_2 \Big\}_{Adj}
 -i\mbox{Tr}_{(4,0)}\ln\bsB_{Adj}
 -i\mbox{Tr}\ln \bm{\nb}^{++}_{Adj}\,.\nn\\ \label{1loop}
  \eea
The first term in \eq{1loop} comes from the quantum gauge multiplet
$v^{++}$ in \eq{Gamma0}. The second term is produced by the
Nielsen--Kallosh ghosts $\chi^{(+4)}$ and $\sigma$. As in the
conventional $\cN=(1,0)$ SYM theory, the last term in \eq{1loop} is
the sum of the contributions coming from the Nielson-Kallosh ghost
$\varphi$ and the Faddeev-Popov ghosts (the analysis of this term
was carried out in \cite{Buchbinder:2016url,Buchbinder:2017ozh}). In
eq. \eq{1loop} the functional trace over harmonic superspace is
defined as
 \be \Tr_{(q,4-q)} {\cal O} = \tr \int d \zeta_1^{(-4)}d
\zeta_2^{(-4)} \, \d_{\cal A}^{(q,4-q)}(2|1)\,  {\cal
O}^{(q,4-q)}(1|2),  \label{trace}
 \ee
where $\d_{\cal A}^{(q,4-q)}(2|1)$ is an
analytic delta-function and ${\cal
O}^{(q,4-q)}(\zeta_1,u_1|\zeta_2,u_2)$ is the kernel of some
operator ${\cal O}$ acting in the space of analytic superfields
possessing  the harmonic $U(1)$ charge $q$ \cite{Galperin:2001uw}.

\section{Power counting and counterterms}
\label{Section_Index}

The general form of possible counterterms can be found using the power
counting in $6D,\, \cN~=~(1,0)$ harmonic superspace. Let us consider
an arbitrary $L$-loop supergraph containing external and internal
lines of the gauge, hypermultiplet and ghost
superfields\footnote{The details of the supergraph technique for the
theory under consideration are discussed in
\cite{Buchbinder:2020tnc}}. The superfield propagators of vector
multiplet, hypermultiplet and ghosts contain the Grassmann
delta-functions \cite{Buchbinder:2020tnc}. This allows us to
represent any loop supergraph as a single integral over anticommuting
variables\footnote{In supersymmetric theories this statement is
closely related to the non-renormalization theorems, see, e.g.,
\cite{Buchbinder:1998qv} }. Taking into account the locality of
divergences, we conclude that each contribution to the effective
action can be presented as an integral over $d^{14}z = d^6x\,
d^8\theta$. Besides, we should take  account of the fact that the quantum
theory is formulated in the framework of the background field
method, which implies that the quantum effective action bears invariance
under the classical background gauge transformations.

Now the power counting, as usual, can be carried out, based on dimensional reasonings.
A contribution to the dimensionless effective action can formally be written as
\begin{equation}\label{Contribution}
\int d^{14}z\, \prod_k du_k\, \Big[\mbox{Momentum integral}\Big]\,
\Big[(D)^{N_D}\Big]\, \Big[\mbox{Superfields}\Big],
\end{equation}
where the superfields correspond to the external lines, and the
symbol $(D)^{N_D}$ denotes the product of $N_D$ spinor covariant derivatives acting
on these external lines (we assume that the external
momenta in this expression are replaced by the relevant
derivatives acting on the corresponding external lines). By
definition, the degree of divergence $\omega$ coincides with the dimension of the momentum integral in units of mass,
\be
\Big[\mbox{Momentum integral}\Big] = m^\omega.
\ee

\noindent
Therefore, it can be found by analyzing the dimensions of various
factors in (\ref{Contribution}).

The dimension of the anticommuting variables is $[\theta] =
m^{-1/2}$, so $[d^{14}z] = m^{-6}\cdot m^4 = m^{-2}$. The harmonic
variables $u_k^\pm $ are dimensionless. The gauge superfield
$\bm{V}^{++}$ (or $v^{++}$) is dimensionless, while the
hypermultiplet and the Faddeev--Popov ghosts have the dimension
$m$. Thus, the dimension of the external legs is $m^{2N_q+2N_c}$, where
$N_q$ and $N_c$ are the numbers of external hypermultiplet and ghost
lines, respectively. If $N_D$ spinor derivatives (each of the
dimension $m^{1/2}$) act on external lines, then this dimension is
increased by $m^{N_D/2}$. Each gauge multiplet propagator present in
the supergraph contributes the factor $g_0^2$.

Taking into account that the effective action is dimensionless, we
obtain $1 = m^{-2}\cdot m^{N_D/2 + 2N_q+2N_c} \cdot m^\omega$, whence
\begin{equation}
\omega = 2 - 2N_q - 2N_c - N_D/2.
\end{equation}

\noindent This implies that the degree of divergence does not depend
on the number of loops and the number of external gauge superfield lines.
Moreover, taking into account that $N_q$ and $N_c$ are even, we see
that divergences can appear only in supergraphs with external gauge
lines. Therefore, all supergraphs with the hypermultiplet or ghost
external lines are finite. Since the theory is formulated in the
framework of the background field method, the form of the
divergences is restricted by the gauge invariance. As a result, it
becomes possible to list all divergences which could present in the theory.

The only gauge invariant combination of the dimension $m^{-2}$
corresponding to the quadratic divergences is proportional to the
standard $6D$, ${\cal N}=(1,0)$ SYM action \cite{Zupnik:1986da}
\begin{equation}\label{N=(1,0)_SYM}
S_{SYM} = \frac{1}{f_0^2} \sum\limits_{n=2}^\infty
\frac{(-i)^{n}}{n} \mbox{tr} \int d^{14}z\, du_1 \ldots du_n\,
\frac{V^{++}(z,u_1)\ldots V^{++}(z,u_n)}{(u_1^+ u_2^+) \ldots (u_n^+
u_1^+)}\,.
\end{equation}
In this case $N_D=0$, so that $\omega=2$.

The only invariant of the dimension $m^0$ corresponding to the
logarithmic divergences reads
\begin{equation}\label{admissible}
\mbox{tr} \int d^{14}z\, du\, V^{--} (D^+)^4 V^{--} = \mbox{tr} \int
d\zeta^{(-4)} du\, \big(F^{++}\big)^2.
\end{equation}
It contains four spinor derivatives acting on the gauge superfields,
so that $N_D=4$ and $\omega=0$ for it.

Therefore, the theory described by the action
\begin{eqnarray}\label{General_Theory}
S = S_{SYM} +\frac{1}{4g_0^2} \int d\zeta^{(-4)} du
\big(F^{++A}\big)^2 - \int d\zeta^{(-4)}\, du\, \widetilde q^{+}
\nabla^{++} q^+
\end{eqnarray}
is renormalizable: all divergences can be absorbed into the
renormalization of the coupling constants $g_0$ and $f_0$. The
hypermultiplet and ghost superfields are not renormalized.\footnote{The Nielson-Kallosh ghosts interact only with the background gauge
superfield and, therefore, no counterterms are required in this sector.}

When using the dimensional regularization (in the context of
superfield theories it is necessary to use its modification called
the dimensional reduction), only the logarithmic divergences are displayed. Therefore, the counterterms of the form (\ref{N=(1,0)_SYM})
are prohibited and the only admissible counterterm is
(\ref{admissible}), {\it i.e.} it is proportional to
the classical action (\ref{S0}) at any loop.

\section{One-loop divergences}
\label{Section_Divergences}

In this section, we will focus on the one-loop quantum correction
$\Delta \Gamma^{(1)}[\bm{V}^{++}]$ to the classical action. We use the
regularization by dimensional reduction and calculate the divergent
part of the one-loop effective action
$\Delta\Gamma^{1}_\infty[\bm{V}^{++}]$ in the lowest order in the
parameter $(\xi_0-1)$ which is assumed to be small.

The most difficult task is to single out the divergent part of the first term in the expression (\ref{1loop})
\begin{eqnarray}\label{Main_Term}
&&\frac{i}{2}\mbox{Tr}_{(2,2)} \ln \bigg\{\Big\{ \frac{1}{\xi_0}
(\bsB_1)^2\, (D_1^+)^4\delta^{(-2,2)}(u_1,u_2)
+ \Big(1-\frac{1}{\xi_0}\Big) \frac{(D_1^+)^4 \bsB_2 (D_2^+)^4}{(u_1^+ u_2^+)^2} e^{i\bm{b}_1} e^{-i\bm{b}_2}
\qquad\nonumber\\
&& + \frac{(D_1^+)^4 (D_2^+)^4}{(u_1^+ u_2^+)^2} e^{i\bm{b}_1} e^{-i\bm{b}_2}
\Big[\, i (\bm{\nabla}^{--}\bm{F}^{++}) -
\frac{i(u_1^+ u_2^-)}{(u_1^+ u_2^+)} \Big(1+\frac{1}{\xi_0}\Big) \bm{F}^{++} \Big]_2 \Big\}_{Adj}\delta^{14}(z_1-z_2)\bigg\},\qquad
\end{eqnarray}
where
$\d^{14}(z_1-z_2)=\d^8(\theta_1-\theta_2)\d^6(x_1-x_2)$.
We present it as the sum of two logarithms,
\begin{eqnarray}\label{Two_Logarithms}
&&
 \frac{i}{2}\mbox{Tr}_{(2,2)} \ln \Big\{\frac{1}{\xi_0}(\bsB_1)^2
 (D_1^+)^4\delta^{(-2,2)}(u_1,u_2)\delta^{14}(z_1-z_2)\Big\} \nn \\
&&
 +\frac{i}{2}\mbox{Tr}_{(2,2)}\ln\bigg\{(D_1^+)^4\delta^{(-2,2)}(u_1,u_2)\delta^{14}(z_1-z_2)
 + \frac{1}{(\bsB_1)_{Adj}^2}\Big\{(\xi_0-1) \frac{(D_1^+)^4 \bsB_2 (D_2^+)^4}{(u_1^+ u_2^+)^2}
 e^{i\bm{b}_1} e^{-i\bm{b}_2} \qquad\nn \\
&&
 + \frac{(D_1^+)^4 (D_2^+)^4}{(u_1^+ u_2^+)^2}
 e^{i\bm{b}_1} e^{-i\bm{b}_2} \Big[\, i\xi_0 (\bm{\nabla}^{--}\bm{F}^{++})
 - (\xi_0+1)\frac{i(u_1^+ u_2^-)}{(u_1^+ u_2^+)} \bm{F}^{++}
 \Big]_2\Big\}_{Adj} \delta^{14}(z_1-z_2)\bigg\}.
\end{eqnarray}
According to \cite{Buchbinder:2017ozh,Buchbinder:2020tnc}, the first
term in this expression vanishes. To calculate the divergent part of
the second term in the lowest order in $(\xi_0-1)$, we need to expand the
logarithm  up to a linear term only. Then we are to find the
divergent part of the expression
\begin{eqnarray}
\Delta\Gamma^{(1)}[\bm{V}^{++}] = \Gamma_1 + \Gamma_2 + \Gamma_3 + \Gamma_4,
\end{eqnarray}
where
\begin{eqnarray}\label{Gamma1}
&& \Gamma_1 = \frac{1}{2} \mbox{tr} \int d\zeta^{(-4)}_1 du_1\, \frac{(\xi_0+1)}{(\bsB_1)^2} (D_1^+)^4 (D_2^+)^4\frac{(u_1^+ u_2^-)}{(u_1^+ u_2^+)^3} e^{i\bm{b}_1} e^{-i\bm{b}_2} \bm{F}^{++}_2
\delta^{14}(z_1-z_2)\Big|_{2\to 1},\qquad\\
\label{Gamma2}
&& \Gamma_2 = -\frac{1}{2} \mbox{tr} \int d\zeta^{(-4)}_1 du_1\, \frac{\xi_0}{(\bsB_1)^2} \frac{(D_1^+)^4 (D_2^+)^4}{(u_1^+ u_2^+)^2} e^{i\bm{b}_1} e^{-i\bm{b}_2} (\bm{\nabla}^{--}\bm{F}^{++})_2\,
\delta^{14}(z_1-z_2)\Big|_{2\to 1},\qquad\\
\label{Gamma3}
&& \Gamma_3 = \frac{i}{2} \mbox{tr} \int d\zeta^{(-4)}_1 du_1\, \frac{(\xi_0-1)}{(\bsB_1)^2} (D_1^+)^4 \bsB_2 (D_2^+)^4 \frac{1}{(u_1^+ u_2^+)^2} e^{i\bm{b}_1} e^{-i\bm{b}_2} \delta^{14}(z_1-z_2)\Big|_{2\to 1},\\
\label{Gamma4}
&& \Gamma_4 = -i\mbox{Tr}\ln \bm{\nb}^{++}\vphantom{\frac{1}{2}},
\end{eqnarray}
and $\mbox{tr}$ stands for the usual matrix trace. The gauge and bridge ($\bm{b}$) superfields appearing in all these expressions
should be expanded over the generators of the adjoint representation. Note that the term
$-i\mbox{Tr}_{(4,0)}\ln\bsB_{Adj}$ present in Eq. (\ref{1loop}) vanishes, see \cite{Buchbinder:2017ozh,Buchbinder:2020tnc}.

Before starting the calculation we outline our strategy. The one-loop effective action \eq{Two_Logarithms} contains the functional trace $\Tr$ defined by Eq. \eqref{trace}. The calculation of this trace includes
the evaluation of the coincident-points limit for the kernel of the corresponding operator. In the expressions \eqref{Gamma1}, \eqref{Gamma2} and \eqref{Gamma3}
the operator $\sB{}^{-2}$ acts on everything to the right of it, including the delta-function.
Hence, before taking the coincident-points limit we should accurately calculate the action of the operator $\sB{}^{-2}$ on all terms to the right
and then single out the divergent contributions.

As the first step of the calculation, we consider the divergent contribution coming from \eq{Gamma1},
 \be
\Gamma_{1}= \f{(\xi_0+1)}{2} \int d\zeta_1^{(-4)} du_1\, ((\bsB_1){}^{-2})^{IJ} (D^+_1)^4 (D^+_2)^4\frac{(u_1^+ u_2^-)}{(u_1^+u_2^+)^3} (e^{i\bm{b}_1} e^{-i\bm{b}_2})^{JK}
(\bm{F}^{++}_2)^{KI} \d^{14}(z_1-z_2)\Big|_{2\to
 1}\,. \label{Gamma1_1}
 \ee
In this expression we should pull out the operator $(\bsB)^{-2}$ to the right. Acting on analytic superfields, the covariant d'Alembertian
\eq{Box_First_Part} yields
$$ \bsB{}^{IJ} =
\partial^2 \d^{IJ} + i(\bm{F}^{++})^{IJ} D^{--} + \dots\,,
$$
where $(\bm{F}^{++})^{IJ} = - if^{KIJ} \bm{F}^{++K}$. The logarithmically divergent
contribution in \eq{Gamma1_1} is proportional to the third inverse
power of the operator $\partial^2 =
\partial^M \partial_M$  acting on the space-time
delta-function $\d^6(x_1-x_2)$. Indeed,
 \bea
 \f{1}{(\pr^2)^3} \delta^6(x_1-x_2)\Big|_{2\to 1} =
 \f{i}{(4\pi)^3 \varepsilon}\,, \quad \varepsilon\to 0\,.
 \label{div}
 \eea
To calculate the coincident-points limit for Grassmann variables, we
use the identity
 \bea
 (D_1^+)^4 (D_2^+)^4 \delta^8(\theta_1-\theta_2) = (u^+_1 u^+_2)^4
 (D_1^+)^4 (D_1^-)^4 \delta^8(\theta_1-\theta_2).
 \label{Id}
 \eea
Then
 \bea
 \Gamma_{1} \to \f{(\xi_0+1)}{2} \int d\zeta_1^{(-4)} du_1\,((\bsB_1)^{-2})^{IJ}
 (u^+_1 u^+_2) (u_1^+ u_2^-) (D_1^+)^4 (D_1^-)^{4}(\bm{F}^{++}_2)^{JI}
 \d^{14}(z_1-z_2)\Big|_{2\to 1}\,.
 \label{Gamma1_2}
 \eea
To evaluate this expression, we note that the derivatives $D^{--}$ inside the operator $\bsB{}^{-2}$ can
act on $(u_1^+ u_2^+)$. Then, taking into account that $D_1^{--}(u^+_1 u^+_2)|_{2\to 1} = - 1$ (see,
e.g.,\cite{Galperin:2001uw}), we obtain that the divergent part of \eq{Gamma1_2} is reduced to
 \bea
\Gamma_{1,\,\infty}= i(\xi_0+1) \int d\zeta^{(-4)}_1 du_1\,(\bm{F}^{++})^{IJ}(\bm{F}^{++})^{JI}
 \f{(D_1^+)^4 (D_1^-)^{4}}{(\partial^2)^3}\d^{14}(z_1-z_2)\Big|_{2\to 1}\,.
 \label{Gamma1_3}
 \eea
Finally we annihilate the Grassmann delta-function, using the identity
$(D_1^+)^{4}(D_1^-)^{4}\d^{8}(\theta_1-\theta_2)|_{2\to 1} = 1$ and
eq. \eq{div}. Then we obtain
 \bea
\Gamma_{1,\,\infty} = - 2(\xi_0+1) \f{C_2}{(4\pi)^3 \varepsilon}
 \tr \int d\zeta^{(-4)} du\, (\bm{F}^{++})^2\,,
 \label{Gamma1_Infty}
 \eea
where $C_2$ is the second Casimir for the adjoint representation of
the gauge group.

The divergent part of the expression (\ref{Gamma2}) vanishes. Indeed, after taking the coincident-points limit we obtain
 \bea
 \Gamma_2 \to -\f{(\xi_0-1)}{4} \int d\zeta^{(-4)}_1 du_1\,((\bsB_1)^{-2})^{IJ}
 (u^+_1 u^+_2)^2 (\bm{\nb}^{--} \bm{F}^{++}_2)^{JI} (D_1^+)^4 (D_1^-)^4
 \d^{14}(z_1-z_2)\Big|_{2\to 1}\,.
 \eea
To annihilate the factor $(u^+_1u^+_2)^2$ we have to expand the
operator $(\bsB{}^2)^{-1}$ up to the second order, so as to gain two derivatives $D^{--}$. As a
result, we accumulate the fourth power of inverse $\pr^2$ operator.
Hence, this term does not contain UV divergence,
\be\label{Gamma2_infty}
\Gamma_{2,\,\infty} = 0.
\ee

One more divergent contribution comes from $\Gamma_3$ given by \eq{Gamma3}. We again use the properties of the spinor derivatives $D^+_a$ and transform the corresponding expression to the form
 \bea
\Gamma_{3} \to \f{i(\xi_0-1)}{2}\int d \zeta^{(-4)}_1 du_1\,
((\bsB_1)^{-2})^{IJ} \bsB_2{}^{JI} (u_1^+u_2^+)^2\, (D_1^+)^4 (D_1^-)^4 \d^{14}(z_1-z_2)\Big|_{2\to 1}\,. \label{Gamma1_5}
 \eea
Also we reconstruct the full superspace measure in \eq{Gamma1_1}
using the property $d^{14}z = d\zeta^{(-4)} (D^{+})^4$. Then the expression \eq{Gamma1_5} can be rewritten as
 \bea
&& \f{i(\xi_0-1)}{4}\int d^{14}z_1 du_1\,
((\bsB_1)^{-2})^{IJ} ((\bm{\nabla}_2^{--})^2)^{JI}\, (u_1^+ u_2^+)^2\, (D_1^+)^4 (D_1^-)^4 \d^{14}(z_1-z_2)\Big|_{2\to 1} \nn\\
 && = \f{i(\xi_0-1)}{4}\int d^{14}z_1 du_1\,
 ((\bm{\nabla}_1^{--})^2)^{JI} ((\bsB_1)^{-2})^{IJ}\, (u_1^+ u_2^+)^2\, (D_1^+)^4 (D_1^-)^4 \d^{14}(z_1-z_2)\Big|_{2\to 1} \qquad\nn\\
 && = \f{i(\xi_0-1)}{2}\int d\zeta^{(-4)}_1 du_1\,
 ((\bsB_1)^{-1})^{II}\, (u_1^+ u_2^+)^2\, (D_1^+)^4 (D_1^-)^4 \d^{14}(z_1-z_2)\Big|_{2\to 1}\,. \qquad\label{Gamma1_6}
 \eea
When acting by the operator $\bsB{}^{-1}$ on the right, we expand the inverse operator $\bsB$ up to the second
order in $D^{--}$ to remove $(u_1^+u_2^+)^2$. After this, we calculate the coincident-points limit
and extract the divergent contribution from \eq{Gamma1_6} as
  \bea\label{Gamma3_infty}
\Gamma_{3,\,\infty}= 2(\xi_0-1) \f{C_2}{(4\pi)^3 \varepsilon} \tr \int d\zeta^{(-4)} du\, (\bm{F}^{++})^2\,.
  \eea

The divergent contribution from $\Gamma_4$ in \eq{Gamma4} was
considered earlier in  \cite{Buchbinder:2016url}. It is
 \bea\label{Gamma4_Infty}
 \Delta\Gamma_{4,\,\infty}= \f{C_2}{3(4\pi)^3 \varepsilon}\,
 \tr\int d\zeta^{(-4)} du\, (\bm{F}^{++})^2\,.
 \eea

Summing up the contributions \eq{Gamma1_Infty}, (\ref{Gamma2_infty}), (\ref{Gamma3_infty}), and (\ref{Gamma4_Infty})
we obtain the final result for the divergent part of the one-loop effective action,
 \be
 \Delta\Gamma^{(1)}_{\rm \infty} = \Gamma_{1,\,\infty} + \Gamma_{2,\,\infty} + \Gamma_{3,\,\infty} + \Gamma_{4,\infty} = -\f{11}{3} \f{C_2}{(4\pi)^3 \varepsilon} \tr
 \int d\zeta^{(-4)} du\, (\bm{F}^{++})^2\,.
 \label{result}
 \ee
We see that all divergent contributions depending on the
gauge-fixing parameter $\xi_0$ in the considered approximation
cancel each other. This agrees with the general statement that the
renormalization of dimensionless coupling constants in
multiplicatively renormalizable gauge theories does not depend on
the gauge choice \cite{Voronov:1982ph,Barnich:2000yx}. Therefore,
the cancelation of terms containing the gauge-fixing parameter can
be considered as a non-trivial test for the correctness of our
calculations. As we have already mentioned, the wave function is not
renormalized in the background (super)field method, so that all
divergences are absorbed into the renormalization of the coupling
constant. The coefficient agrees with the one obtained earlier in
the Feynman gauge $\xi_0=1$, both in the component approach
\cite{Ivanov:2005qf,Casarin:2019aqw} and by the supergraph technique
\cite{Buchbinder:2020tnc}.

\section{Adding the hypermultiplet}
\label{Section_Hypermultiplet}

In this section we consider the one-loop divergences for the theory with the action \eq{S02}, focusing on the divergences
in the hypermultiplet sector. To study possible divergent contributions to the effective action
we introduce the background-quantum splitting for both $q^{+}$ and $V^{++}$,
 \bea
V^{++} \to \bm{V}^{++} + v^{++}\,, \qquad  q^+ \to Q^{+} + q^{+}\,.
 \eea
Here we have denoted the ``background'' superfields by the capital
letters $\bm{V}^{++}, \, Q^{+}$  and ``quantum'' ones by $v^{++}, \,
q^+$. The presence of the background hypermultiplets leads to the mixing of the superfields $v^{++}$ and $q^{+}$. All such terms can
be eliminated by a special redefinition of the quantum hypermultiplet in the functional integral similarly to the case of $6D$, $\cN=(1,0)$
SYM theory (see \cite{Buchbinder:2016url} for details). The calculation of the one-loop divergences for the theory \eq{S02}
is performed in a close analogy to the case of the model
\eq{S0} described in Sec.5. The effective action is written as
   \bea
 &&\hspace*{-7mm} \Delta\Gamma^{(1)}[\bm{V}^{++},Q^+] =
 \frac{i}{2}\, \mbox{Tr}_{(2,2)}\ln \Big\{\Big\{ \frac{1}{\xi_0}
(\bsB_1)^2\, (D_2^+)^4\delta^{(2,-2)}(u_1,u_2) +
\Big(1-\frac{1}{\xi_0}\Big) \frac{(D_1^+)^4 \bsB_2 (D_2^+)^4}{(u_1^+
u_2^+)^2} e^{i\bm{b}_1} e^{-i\bm{b}_2}
\nonumber\\
&&\hspace*{-7mm} + \frac{(D_1^+)^4 (D_2^+)^4}{(u_1^+ u_2^+)^2} e^{i\bm{b}_1} e^{-i\bm{b}_2}
\Big[\, i (\bm{\nabla}^{--}\bm{F}^{++}) -
\frac{i(u_1^+ u_2^-)}{(u_1^+ u_2^+)} \Big(1+\frac{1}{\xi_0}\Big) \bm{F}^{++} \Big]_2 \Big\}_{Adj}^{IJ}
 - 2g_0\widetilde Q^{+}_1 T^I G^{(1,1)}(1|2) T^J Q^+_{2}\Big\}
  \nn \\
 &&\hspace*{-7mm} -i\, \mbox{Tr}_{(4,0)}\ln\bsB_{Adj} -i\, \mbox{Tr}\ln \bm{\nb}^{++}_{Adj}
 + i\, \Tr\ln \bm{\nb}^{++}_R\,, \label{1loop3}
 \eea
where
 \be
 G^{(1,1)}(1|2)=\f{1}{\bsB}\frac{(D^+_1)^4(D^+_2)^4}{(u^+_1u^+_2)^3}\, \delta^{14}(z_1-z_2)
 \ee
is the hypermultiplet Green function \cite{Galperin:2001uw} and
$T^I, \, I=1,..,{\rm dim} \,G, $ are generators of
gauge group $G$ in the hypemultiplet representation  $R$.

The term $i\, \Tr\ln \bm{\nb}^{++}_R$ in the expression \eq{1loop3} corresponds to the contribution of the quantum hypermultiplet.
Taking into account that the supergraphs with $Q^+$ on external legs are finite, we see that the hypermultiplet
can merely change the coefficient of the purely gauge contribution containing $\mbox{tr} (\bm{F}^{++})^2$. This implies that
\bea
\Delta\widetilde\Gamma^{(1)}_\infty[\bm{V}^{++}] = \Big(\Delta\Gamma^{(1)}[\bm{V}^{++}] + i\, \Tr\ln \bm{\nb}^{++}_R\Big)_\infty, \label{1loop2}
 \eea
where $\Delta\Gamma^{(1)}[\bm{V}^{++}]$ was introduced earlier in \eq{1loop}, and $\widetilde\Gamma$ is the effective action for the model \eq{S02}.

The contribution of the hypermultiplet to the one-loop divergences has been calculated in \cite{Buchbinder:2016url,Buchbinder:2017ozh}.
Adding it to the expression (\ref{result}), we obtain the total contribution in the form
 \bea
 \widetilde\Gamma^{(1)}_{\rm \infty}[\bm{V}^{++}] = -\f{11C_2+T_R}{3(4\pi)^3 \varepsilon} \tr  \int d\zeta^{(-4)} du\, (\bm{F}^{++})^2\,,
 \label{result2}
 \eea
where the constant $T_R$ is defined by the relation $\tr( T^I T^J) = T_R\, \delta^{IJ}$. Thus, the presence of the hypermultiplet
gives rise to an increase of the absolute value of the $\beta$-function.

Now, let us explicitly verify that the part of the effective action
containing the hypermultiplet is finite\footnote{For simplicity, we omit all gauge group indices and
explicit dependence on the generators.}. To this end, we consider
that term in \eq{1loop3} which depends on the background
hypermultiplet $Q^+$. The corresponding contribution with the
maximal degree of divergence reads \footnote{Other contributions can
be analyzed in a similar manner.}
 \bea
&& \int d\zeta^{(-4)}_1 d\zeta^{(-4)}_2  du_1\, du_2\,
 \bigg(\frac{1}{\xi_0} (\bsB_1)^2 (D_2^+)^4\delta^{(2,-2)}(u_1,u_2)
 +\f{(D^+_1)^4 \bsB_2 (D_2^+)^4}{(u_1^+u_2^+)^2}\Big(1-\frac{1}{\xi_0}\Big)\bigg)^{-1} \nn\\
 && \times \delta^{14}(z_1-z_2) \widetilde Q^{+}_1
G^{(1,1)}(1|2) Q^+_{2} \sim  \int d\zeta_1^{(-4)} du_1\,
\f{\xi_0}{(\bsB_1)^2} \widetilde Q^{+}_1 G^{(1,1)}(1|2)
Q^+_{2}\Big|_{2\to 1} +\ldots, \qquad\label{1loop6}
 \eea
where dots denote terms containing higher powers of $1/\partial^2$. The explicitly written term in this expression can be worked out as
 \bea
&& \int d\zeta_1^{(-4)} du_1\, \f{\xi_0}{(\bsB_{1})^2} \widetilde
Q^{+}_1 G^{(1,1)}(1|2)
 Q^+_{2}\Big|_{2\to 1}\nonumber\\
 && = \int d\zeta_1^{(-4)} du_1\, \f{\xi_0}{(\bsB_{1})^2} \widetilde Q^{+}_1
 Q^+_{2}
 \frac{1}{\bsB_1} (D^+_1)^4(D^-_1)^4 (u_1^+u^+_2) \delta^{14}(z_1-z_2)\Big|_{2\to
 1}, \nn \\
 && \sim \int d\zeta^{(-4)} du\, \widetilde Q^{+} \bm{F}^{++} Q^{+} \,
 \f{1}{(\partial^2)^4} \delta^6(x-x')\big|_{x'\to x}\,, \label{1loop4}
 \eea
where we have used the identity \eq{Id}. In the expression \eq{1loop4}, using the relation similar to \eq{Box_First_Part},  we
expand  the operators $\bsB$  up to the first order in $\bm{F}^{++}\nb^{--}$ and then act by the harmonic derivative
$\bm{\nb}^{--}$ on the factor $(u_1^+u^+_2)$ in the coincident harmonic points
limit. But the hypermultiplet Green function $G^{(1,1)}$ brings the
inverse power of operator $\bsB$ and the resulting power of
$\partial^2$ in the denominator amounts to a finite contribution to
the effective action. The remaining terms in \eq{1loop6} contain at least the fourth power of $\bsB$ in the
denominator. Therefore, they are also finite, in agreement with the power counting arguments of section \ref{Section_Index}.
This means that the one-loop divergencies in the hypermultiplet sector are actually absent.

\section{Summary and outlook}
\label{Section_Conclussion}

We studied the quantum divergence structure of the higher-derivative
$\cN=(1,0)$ supersymmetric non-abelian gauge theory in six dimensions.
This theory involves four derivatives in the component gauge field  sector and
three derivatives in the spinor gaugino sector. The theory is
characterized by a dimensionless coupling constant. Two such models
were considered: the model with the gauge multiplet only and the
model in which the gauge multiplet is coupled to the hypermultiplet  in some
representation of the gauge group, with the standard kinetic terms for
the hypermultiplet physical scalars and fermions.
Both models were formulated in harmonic $6D, \,\cN=(1,0)$ superspace ensuring manifest $\cN=(1,0)$
supersymmetry. The quantization was accomplished in the framework of
the background superfield method with a one-parameter family of the quantum gauge-fixing conditions.
The corresponding gauge invariant and manifestly supersymmetric
quantum effective action was introduced and all possible divergent terms in such an action were identified.

The analysis of the divergence structure for the theories under consideration was based on
the superfield power counting. It was shown
that the superficial degree of divergence does not depend on a number of loops and
is completely specified by the number of $D$-factors acting on the
external lines of the gauge superfield. Then, taking into account the gauge invariance
of the effective action and making use of the regularization by dimensional reduction, we conclude that the only possible
counterterm in the theory is proportional to the  classical action of gauge superfield.
All supergraphs with the hypermultiplet external legs should be finite.
This implies that the theory under consideration is multiplicatively
renormalizable and the renormalization affects only the dimensionless
coupling constant.

A manifestly supersymmetric and gauge invariant procedure to
calculate the one-loop divergences was developed and applied for the explicit calculation
of these divergencies. The result completely agrees with the one obtained earlier in \cite{Ivanov:2005qf, Casarin:2019aqw,Buchbinder:2020tnc}
through direct calculations of Feynman (super)graphs, as well as with the general
analysis based on the power counting.  It was also shown
that in the lowest order with respect to the deviation  $(\xi_0-1)$ of the gauge-fixing parameter $\xi_0$  from
its minimal-gauge value $\xi_0 = 1$ the divergences are independent of this parameter, which can be
considered as a check of the correctness of our calculations. We also found the modification of the one-loop
divergence in the gauge superfield sector by the hypermultiplet contribution. It amounts to changing the absolute value of
the relevant coefficient.

There are at least four interesting directions for further generalization
of the results obtained.
\begin{itemize}
\item{It would be tempting to calculate the one-loop divergencies for the
general theory (\ref{General_Theory}) the action of which is a sum of the higher-derivative action (\ref{S0})
and the action of the standard $6D, \, \cN=(1,0)$ SYM theory. In this case
we will deal with one dimensionless coupling constant $g_0$ and another
dimensionful coupling constant $f_0$. Such a theory is still
multiplicatively renormalizable, but there can be non-trivial running coupling constant regimes.}
\item{It is interesting to develop the superfield method for studying the superconformal anomaly
in the higher-derivative theory (\ref{S0}).}
\item{One more noteworthy prospect is to study a renormalization structure of the higher-derivative gauge
superfield model coupled to the higher-derivative hypermultiplet model. The corresponding classical theory
was constructed in \cite{Ivanov:2005kz}. In principle, such a
consideration could allow to set up $6D, \,\cN=(1,1)$ supersymmetric anomaly-free
higher-derivative theory. One can, e.g.,  conjecture that this theory is asymptotically free and even completely finite.}
\item{An obvious generalization of our study is to find the total dependence of the divergences on the gauge-fixing parameter $\xi_0$.}

\end{itemize}
We hope to address all these problems in the forthcoming works.

\section*{Acknowledgements}
The authors are grateful to Andrei Smilga for valuable comments. The
work was supported by the grant of Russian Science Foundation,
project No. 16-12-10306.

\end{document}